\documentclass[twocolumn,aps,prl,showpacs]{revtex4}
\usepackage{graphicx}
\usepackage{epsf}
\newcommand{\etal}{{\it et al.}}

\begin{document}

\title{How to measure a spinon Fermi surface}

\author{M. R. Norman and T. Micklitz}
\affiliation{Materials Science Division, Argonne National Laboratory, Argonne, IL 60439}

\begin{abstract}
We propose an experiment to identify the potential existence of a spinon Fermi surface
by looking for oscillatory coupling between two ferromagnets via a spin liquid spacer.
Three candidate spin liquids are investigated, and it is found that in all cases,
long period oscillations should be present, the period of which would identify the
Fermi wavevector of the spinon surface.
\end{abstract}
\date{\today}
\maketitle

In 1973, Anderson~\cite{phil73} proposed the possibility of a spin liquid, a state where long range
magnetic order is suppressed by frustration, low dimensionality and/or quantum fluctuations.
In 1987, he resurrected this idea in the context of high temperature cuprate 
superconductors~\cite{phil87}.  The proposed ground state was a so-called uniform resonating
valence bond (RVB) state which possesses a Fermi surface for spin excitations.  Although it
turns out that the undoped cuprates are antiferromagnets, the spin liquid concept in the context
of doped cuprates is a very active field of study~\cite{leeRMP}.

Frustration plays a major role in suppressing magnetic order, and the original 
RVB idea was developed for a triangular lattice.  Since then, other frustrated lattices have been
identified, including pyrochlore, kagome, and hyper-kagome~\cite{ramirez}.  In the past few years,
three candidate $S=1/2$ spin liquids have been identified~\cite{leeNV}:
$\kappa$-(BEDT-TTF)$_2$Cu$_2$(CN)$_3$ (distorted triangular lattice)~\cite{shimizu},
ZnCu$_3$(OH)$_6$Cl$_2$ (kagome)~\cite{shores},
and Na$_4$Ir$_3$O$_8$ (hyper-kagome)~\cite{okamoto}.
All of these materials are insulating, have a large Curie-Weiss temperature, and yet show
no ordering down to the lowest temperatures measured.  In each case~\cite{motrunich, hermele,
ma,zhou,lawler}, a uniform RVB state has been proposed as the ground state with a resulting spinon
Fermi surface.  There is indirect evidence for such a surface.  The BEDT salt exhibits a linear $T$
specific heat~\cite{yamashita} characteristic of a Fermi surface.  The herbertsmithite
ZnCu$_3$(OH)$_6$Cl$_2$
exhibits a sub-linear $T$ dependence of the specific heat~\cite{helton},
but this can be understood from self-energy 
corrections~\cite{motrunich}. In addition, a Pauli-like susceptibility is inferred at low
temperatures once the effect
of impurity spins has been factored out.

On the other hand, this evidence for a spinon surface is indirect.
Spin glass behavior, or a magnon dispersion, $\omega^z$,
with a power $z$ equal to the dimensionality, can also lead to a linear $T$ specific heat~\cite{ramirez}.
Therefore, it would be desirable to have a more direct test.  Now, the existence of a Fermi surface
implies the presence of Friedel oscillations.  For a spin liquid, this would not show up in the charge
channel, but would show up in the spin channel.  The challenge is how to detect this.

As is well known, oscillatory coupling has been seen between two ferromagnets separated by a
paramagnetic spacer (Fig.~1)~\cite{parkin}.
This is a consequence of Kohn anomalies of the Fermi surface
which appear in the spin susceptibility~\cite{bruno},
but are difficult to observe by neutron scattering because of their weak intensity.
%MRN - changes
These anomalies are due to extremal spanning vectors of the Fermi surface.
There are several types:
2$k_F$ vectors, umklapp vectors, and vectors which connect different Fermi surfaces.
Large vectors are difficult to observe in the oscillatory coupling since their period is comparable 
to the lattice constant~\cite{motrunich} (and are therefore damped by roughness of the layers), but
small vectors have been prominently observed in transition metal multi-layers~\cite{parkin}.
The existence of small vectors (long period oscillations) is generic, as large Fermi
surfaces have small umklapp vectors, whereas small Fermi surfaces have small 2$k_F$ vectors.
Oscillatory experiments have the additional advantage of being able to detect multiple $q$ vectors
with different orientations depending on the growth direction of the layers.
These oscillations are also strongly suppressed by a spin gap~\cite{bruno2}, as the latter leads to 
an exponential decay of the susceptibility with a distance scale set by the lattice constant~\cite{jiang}.

\begin{figure}
\centerline{\includegraphics[width=2.0in]{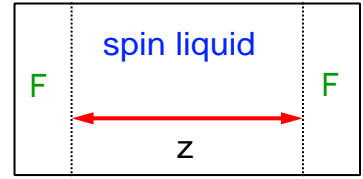}}
\caption{(Color online) The proposed experiment involves two ferromagnetic layers (F) with a 
spin liquid spacer
of variable thickness, $z$.
Depending on the sign of the oscillatory coupling, the two ferromagnets will be aligned
or anti-aligned.}
\label{fig1}
\end{figure}

We will now apply this idea to spin liquids.  We start with the BEDT salt, originally discovered
by Geiser \etal~\cite{geiser}.  As discussed by
Shimizu \etal~\cite{shimizu}, this material is composed of dimers that sit on a distorted triangular
lattice.  This particular salt, though, has a very small anisotropy of the hopping integrals of 6\%.
The resulting Fermi surface is shown in Fig.~2a, which was derived from the dispersion
$\epsilon_k = 2t'\cos(k_bb)+4t\cos(k_bb/2)\cos(k_cc/2)-\mu$ with $t$=54.5 meV, $t'$=57.5 meV,
and $\mu$=-46.2 meV to achieve half-filling ($b$ and $c$ are the orthorhombic lattice constants 
parallel to the layers).  This surface is removed by a Mott
transition~\cite{kurosaki}, but the transition is suppressed by pressure.  We assume that
the spinon surface in the Mott phase is the same as the `band' Fermi surface~\cite{foot1}.
We note that the shortest
spanning vector along $k_c$ is an umklapp vector of length 0.94$\pi/c$=0.22$\AA^{-1}$, which 
would give 
rise to a real space period of 28.5$\AA$, comparable to that seen in multi-layers~\cite{parkin}.
We should remark that the real Fermi surface (under pressure) has been seen both by
angle dependent magnetoresistance and Shubnikov-de Haas oscillations~\cite{ohmichi}.

We can perform the same exercise for the herbertsmithite~\cite{shores}.
This material has copper atoms on
a kagome lattice, with three copper atoms per unit cell.
Assuming a near-neighbor spinon hopping, $t_s$, one has three bands, the
middle of which is half-filled.  Its
dispersion is given by $E/t_s = 1 - X - \mu$,
where $4X^2 = t_{12}^2+t_{13}^2+t_{23}^2+3t_{12}t_{13}t_{23}$
with $t_{12}=2\cos(k_ya/2), t_{13}=2\cos(k_ya/4+\sqrt{3}k_xa/4), t_{23}=2\cos(k_ya/4-\sqrt{3}k_xa/4)$
and $a$ the lattice constant (each bond has a length $a/2$).  The spinon Fermi surface,
previously derived by Ma and Marston~\cite{ma}, is shown in Fig.~2b.  It has a short umklapp spanning 
vector of 0.50$(2\pi/\sqrt{3}a)$=0.265$\AA^{-1}$, giving rise to a period of 23.7$\AA$, similar to that
estimated for the BEDT salt.

\begin{figure}
\centerline{\includegraphics[width=3.4in]{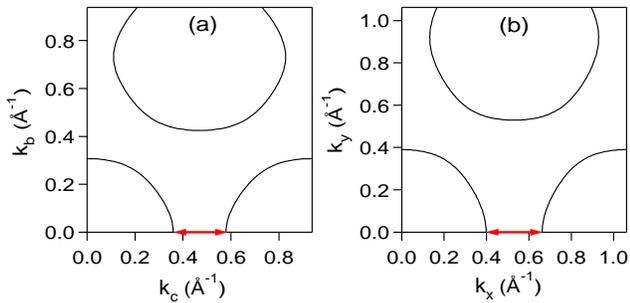}}
\caption{(Color online) Spinon Fermi surface for 
(a) $\kappa$-(BEDT-TTF)$_2$Cu$_2$(CN)$_3$ ($b$=8.59$\AA$,
$c$=13.40$\AA$)
and (b)  ZnCu$_3$(OH)$_6$Cl$_2$ ($a$=6.84$\AA$).  Spanning vectors are indicated by arrows.}
\label{fig2}
\end{figure}

One issue with these two materials is that they are two dimensional.  To see an oscillatory
period as discussed above requires that the spanning vector have a component perpendicular
to the multi-layer (i.e., along $z$ in Fig.~1).
This would require growing the material with a surface not perpendicular to
the `(001)' direction, which may be a difficult undertaking.  Along with this would be 
the probable difficulty of growing these materials with controlled thickness between 
the two ferromagnets.

We now turn to the hyper-kagome case.   Na$_4$Ir$_3$O$_8$ has several advantages.  It is
cubic, and is also a transition metal oxide.  As such, it should not have some of the growth
difficulties mentioned
above.  This material has 12 Ir atoms in the unit cell~\cite{okamoto}.
Each Ir atom has 4 Ir neighbors, forming a network of
corner sharing triangles.  Assuming again near-neighbor hopping,
the resulting 12 bands can easily be found by numerical
diagonalization.  The spinon Fermi surface for this case has been shown by 
Zhou \etal~\cite{zhou}, and consists of two hole pockets around the $R$ point of the simple
cubic zone, and one electron pocket around the $\Gamma$ point (Fig.~3a).  The surfaces are
small, and thus the shortest spanning vectors are of the $2k_F$ variety.  To a first approximation,
we find that
the surfaces can be approximated by spheres (the cubic anisotropy is of order 10\%).
The $k_F$ vector for the electron surface along (100) is 0.31$\pi/a$ and that of the hole pockets
0.26$\pi/a$.  As there are two hole pockets, their contribution will dominate over the 
electron pocket in the oscillatory coupling.  With $a$ of 8.985$\AA$, this results in a predicted
oscillatory period of 34.5$\AA$.

%MRN - addition
We can make a quantitative estimate of the 
oscillatory coupling following the literature in the GMR (giant magnetoresistance) field~\cite{bruno}.
We assume that the coupling of the spins in the ferromagnet to those in the spin liquid can be
described by an effective contact interaction $A \delta({\bf r}-{\bf r}_i) {\bf s} \cdot {\bf S}_i$ 
where ${\bf s}$ is the
spin vector of the spin liquid at position ${\bf r}$ and ${\bf S}_i$ the spin vector of the ferromagnet at
position ${\bf r}_i$.  Values for $A$ can be estimated by calculating the transmission properties of
the barrier between the ferromagnet and spacer layers, although these calculations are quite
involved~\cite{stiles,bruno2}.  This effective contact interaction then leads to a coupling between
a site in the first ferromagnet layer to one in the second
\begin{equation}
J_{ij} = -\frac{A^2}{16\pi^3V} \int d^3{\bf q} \chi({\bf q}) e^{i{\bf q} \cdot ({\bf r}_j-{\bf r}_i)}
\end{equation}
where $V$ is the unit cell volume of the spin liquid, and $\chi({\bf q})$ the static susceptibility
of the spin liquid in units of $2\mu_B^2$ per unit cell.
The interlayer coupling per unit area is then obtained by 
summing over the lattice sites in the second ferromagnetic layer,
\begin{equation}
I_i = \frac{d}{V} S^2 \sum_j J_{ij}
\end{equation}
where $d$ is the layer spacing of the spin liquid.  Substituting for $J_{ij}$, we have
\begin{equation}
I(z) = -\frac{A^2S^2d}{16\pi^3V^2} \int dq_z e^{iq_zz} \int d^2{\bf q}_{\|} \chi({\bf q}_{\|},q_z)
\sum_{{\bf r}_{\|}} e^{i{\bf q}_{\|} \cdot {\bf r}_{\|}}
\end{equation}
with $z$ oriented as in Fig.~1 
and the sum is over $r$ vectors in the second ferromagnetic layer.
By Bloch's theorem, the sum reduces ${\bf q}_{\|}$ to zero, and we obtain
\begin{equation}
I(z) = -\frac{A^2S^2d^2}{4\pi V^3} \int dq_z e^{iq_zz}\chi(0,0,q_z)
\end{equation}

To proceed further, we have to specify $\chi$.  As in the spin liquid literature~\cite{motrunich},
we will assume that $\chi$ is given by the bare polarization bubble times the Gutzwiller projection
factor $g_J = 4$~\cite{footm}.  This then yields
\begin{equation}
I(z) = -\frac{A^2S^2g_Jg^2s^2d^2}{32\pi^4V^2} \int d^3{\bf k} \int dq_z e^{iq_zz}
\frac{f(\epsilon_{\bf k})-f(\epsilon_{\bf k+q})}{\epsilon_{\bf k+q}-\epsilon_{\bf k}}
\end{equation}
where ${\bf q} \equiv (0,0,q_z)$, $f$ is the Fermi-Dirac function, $\epsilon_{\bf k}$ the spinon
dispersion, and $g$ the effective g-factor of the spin liquid
($S$ is the magnitude of the spin vector of the ferromagnet, and $s$ the magnitude of the
spin vector of the spin liquid).  Note that there is an implicit double sum over band indices.

The integrals over $k_z$ and $q_z$ can be converted to integrals over $\epsilon_{\bf k}$
and $\epsilon_{\bf k+q}$, and the remaining integral over ${\bf k}_{\|}$ will be dominated by extremal
$q$ vectors connecting the spinon Fermi surface(s).
The variation of the extremal $q$ vector is given by an expansion of the spinon dispersion
about $k_F$~\cite{koelling}
\begin{equation}
\epsilon_{\bf k} = v_z(k_z-k_F) + \frac{1}{2}(k_x^2D_{xx}+k_y^2D_{yy}+2k_xk_yD_{xy})
\end{equation}
where $v_z$ is the Fermi velocity along $z$ and $D$ is the inverse mass tensor.
In our cubic case, $D_{xx}=D_{yy}$.
Setting $\epsilon_{\bf k}$ to zero and solving, we find
\begin{equation}
k_z = k_F - \frac{1}{2v_z}((k_x^2+k_y^2)D_{xx}+2k_xk_yD_{xy})
\end{equation}
This is easily diagonalized by rotating 45 degrees in $k_x,k_y$ space
\begin{equation}
k_z = k_F - \frac{1}{2v_z}(k_1^2D_1+k_2^2D_2)
\end{equation}
where $D_{1,2}=D_{xx} \pm D_{xy}$.

Because of the multiple Fermi surfaces in Fig.~3a, both intraband and interband terms are present.
The latter will be particularly relevant for the susceptibility in the presence of spin-orbit coupling 
(known to be significant in the case of Na$_4$Ir$_3$O$_8$~\cite{balents}).
For the intraband terms, $q$ is simply twice $k_z$.
Collecting these terms, we arrive at the T=0 expression~\cite{bruno}
\begin{equation}
I(z) = -I_0 \frac{d^2}{z^2} \sum_n \frac{m^*_n}{m} \sin(2k_{Fn}z + \phi_n)
\end{equation}
where $m^*_n$ is the effective mass,
\begin{equation}
m_n^* = 1/\sqrt{|D_{xx}^2-D_{xy}^2|}
\end{equation}
$m$ the bare mass, and $n$ the band index.  Restoring $\hbar$,
$I_0$ is equal to $m g_J (ASgs/(4\pi \hbar V))^2$~\cite{grolier}.
$\phi_n$ is a phase angle, equal to zero if $q$ is a maximum, $\pi/2$ if a saddlepoint,
and $\pi$ if a minimum.
For the interband terms, we must replace $D_1$ and $D_2$ by their averages over the two bands.
For the hyper-kagome case, it turns out that $D_1$ and $D_2$ are interchanged between the
two hole bands.  As a consequence, the interband contributions in this case reduce to the
same expression as in Eq.~9, but with
\begin{equation}
m_n^* = 1/|D_{xx}|
\end{equation}
The various parameters are listed in Table I. 

\begin{table}
\caption{Properties of the hole and electron surfaces of Na$_4$Ir$_3$O$_8$ used in
calculating the oscillatory coupling.  $k$ is in units of $\pi/a$, $m^*$ in units of $m$,
and $\epsilon$ in units of $t_s$.}
\begin{ruledtabular}
\begin{tabular}{cccccc}
band & $k_F$ & $v_z$ & $D_{xx}$ & $D_{xy}$ & $m_n^*$ \\
\colrule
5 & (1,1,1.260) & -0.971 & -4.182 & -2.465 & 0.296 \\
6 & (1,1,1.260) & -0.971 & -4.182 & 2.465 & 0.296 \\
7 & (0,0,0.312) & 1.522 & 2.618 & -0.001 & 0.382 \\
\end{tabular}
\end{ruledtabular}
\end{table}

Note that the value of $m_n^*$ for the interband term from Table I is 0.239.
Summing over the two intraband and two interband terms,
we find the total hole contribution to be twice 0.296 + 0.239 or 1.07.
The ratio of this to the electron value of 0.382 is 2.8.
Noting that in the present case, for all contributions, 
$q$ is a maximum~\cite{foot2}, we then find
\begin{equation}
I(z) = -I_0\frac{d^2}{z^2} (2.8\sin(0.182z)+\sin(0.218z))
\end{equation}
where we have absorbed the proportionality
constant for the mass (which involves the spinon hopping integral $t_s$) into $I_0$,
with $z$ (the separation of the two ferromagnets) in units of $\AA$.
This function is plotted in Fig.~3b, where we find a periodicity
consistent with the hole period of 34.5$\AA$ mentioned above, but with beating
clearly present due to the electron period.  This shows the power of this technique
to resolve complex Fermi surface topologies.

\begin{figure}
\centerline{\includegraphics[width=3.4in]{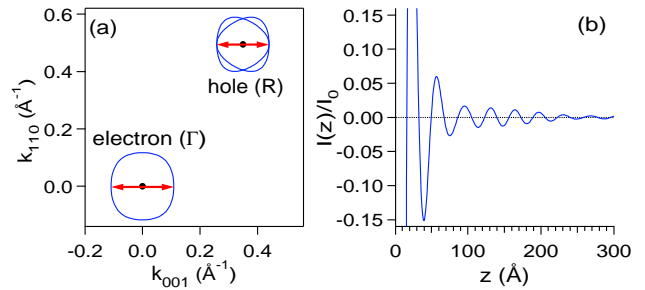}}
\caption{(Color online) (a) Spinon Fermi surface for 
Na$_4$Ir$_3$O$_8$ ($a$=8.985$\AA$).
Spanning vectors are indicated by arrows.
(b) Calculated oscillatory response from Eq.~12.}
\label{fig3}
\end{figure}

We note that although at very low temperatures, Na$_4$Ir$_3$O$_8$ exhibits a super-linear 
specific heat,
indicating that the spinon surface might develop nodes~\cite{zhou}, over most of the temperature
range, it is consistent with an ungapped surface~\cite{lawler}.  Therefore, we regard this cubic material
as a promising one to consider in the context of our proposed experiment, though we encourage
that all candidate spin liquids be looked at.

In conclusion, we have proposed an experiment based on oscillatory coupling between two
ferromagnets with a spin liquid spacer for possible detection of a spinon Fermi
surface.  This, and other Friedel like experiments, will hopefully be pursued in the future to see
whether this novel spin liquid state exists.

Work at Argonne National Laboratory was supported by the U.S. DOE, Office of Science, under Contract 
No.~DE-AC02-06CH11357.  This project was inspired by a talk given by Leon Balents at the ICTP
in Trieste.


\begin{thebibliography}{99}
\bibitem{foot1} The spinon hopping integral, $t_s$, is proportional to the superexchange 
integral, $J$, which goes as the square of the orbital hopping
integral, $t$.  But as the $t$ anisotropy is weak, we will consider the $J$ anisotropy (yet to
be determined experimentally) to be the same as that of $t$.
\bibitem{foot2}  Note that for both cases in Fig.~2, $q$ is a minimum, so the oscillatory
coupling will change sign.
\end{thebibliography}
\end{document}